\documentstyle[preprint,aps,epsfig]{revtex}

\begin{document}

\draft        

\preprint{\font\fortssbx=cmssbx10 scaled \magstep2
\hfill$\vcenter{\hbox{\bf CERN-TH/96-22}
                 \hbox{\bf IFUSP-P 1198}
                 \hbox{\bf MAD-PH-96-926}
                 \hbox{\bf hep-ph/9602264} }$}

\title{Constraints on Quartic Vector-Boson Interactions\\
  from $Z$ Physics }

\author{A.\ Brunstein \cite{ad} }

\address{Instituto de F\'{\i}sica,
Universidade de S\~ao Paulo, \\
C.P.\ 66318, 05389-970 S\~ao Paulo, Brazil}

\author{O.\ J.\ P.\ \'Eboli \cite{ojpe} }

\address{Physics Department, University of Wisconsin, 
Madison, WI 53706, USA}

\author{M.\ C.\ Gonzalez-Garcia \cite{mcgg}}

\address{ CERN, Theory Division, 
CH-1211 Gen\`eve 23, Switzerland}

\date{\today}

\maketitle
\begin{abstract}
\vskip -0.5cm
\baselineskip 0.45 cm
We obtain the constraints on possible anomalous quartic vector-boson
vertices arising from the precision measurements at the $Z$ pole. In
the framework of $SU(2)_L \otimes U(1)_Y$ chiral Lagrangians, we
examine all effective operators of order $D=4$ that lead to four-gauge-boson
interactions but do not induce anomalous trilinear vertices.  We
constrain the anomalous quartic interactions by evaluating their
one-loop corrections to the $Z$ pole physics. Our analysis is
performed in a generic $R_\xi$ gauge and it shows that only the
operators that break the $SU(2)_C$ custodial symmetry get limits close
to the theoretical expectations. Our results also indicate that these
anomalous couplings are already out of reach of the Next Linear $e^+
e^-$ Collider, while the Large Hadron Collider could be able to further
extend the bounds on some of these couplings.
\end{abstract}
\noindent
{\bf CERN-TH/96-22} \\
\noindent
\vskip -1. cm
{\bf February 1996}
\newpage

\section{Introduction}
\label{sec:int}

The standard model (SM) of electroweak interactions has been the
subject of an intense experimental research that confirmed its
predictions for the interactions between fermions and vector bosons
\cite{rolandi}. However, some elements of the SM, such as the symmetry
breaking mechanism and the interaction among the gauge bosons, have
not been the object of direct experimental observation yet.  In
particular, the structure of the triple and quartic vector-boson
couplings is completely determined by the $SU(2)_L \otimes U(1)_Y$
gauge structure of the model, and a detailed study of these
interactions can either further confirm the local gauge invariance of
the theory or indicate the existence of new physics beyond the SM.

Presently only hadronic colliders have directly studied the triple
vertex $W^+W^-\gamma$ \cite{present}; however, their constraints on
this coupling are very loose.  One of the main goals of LEP II at CERN
will be the investigation of the reaction $e^+e^- \rightarrow W^+W^-$,
which can furnish direct bounds on anomalous $W^+W^-\gamma$ and
$W^+W^-Z$ interactions \cite{Dieter1}. Future hadron \cite{anohad} and
$e^+e^-$, $e\gamma$, and $\gamma\gamma$ \cite{anoNLC} colliders will
also provide information on these couplings and improve significantly
our knowledge of possible anomalous gauge-boson interactions.

Direct studies of quartic vector-boson interactions cannot be
performed at the colliders in operation since the available
centre-of-mass energy is not sufficient for multiple vector-boson
production.  This crucial test of the gauge structure of the SM will
be possible only at the CERN Large Hadron Collider (LHC) through the
reaction $pp \rightarrow V_L V_L X$ \cite{deh,bdv,dhpru} or at the
next linear collider (NLC) through the processes $e^+ e^-
\rightarrow VVV$ \cite{bela1,stir}, $e^- e^-\rightarrow F F V V $ \cite{cuypers}, $e\gamma \rightarrow V V F$ \cite{our:eg},
$\gamma\gamma \rightarrow VV$ \cite{bela2}, and $\gamma \gamma
\rightarrow VVV $ \cite{our:vvv}, where $V=$ $Z$, $W^\pm$ or $\gamma$
and $F=$ $e$ or $\nu_e$. However, these machines will not operate in the
near future, and consequently we will have to rely on indirect
information on the quartic vertices for quite some time.

Valuable information on anomalous interactions can also be gathered
from the low energy data \cite{low} and the results of the $Z$ physics
\cite{DeRujula,outros,Dieter2,epsib}, which can also constrain
substantially the possible deviations of the gauge boson
self-interactions from the SM predictions through their contributions
to the electroweak radiative corrections. So far all the analyses have
concentrated on operators which generate tree-level modifications to
the gauge-boson two-point or three-point functions.

In this work, we obtain the constraints on quartic vector-boson
self-interactions arising from the precision measurements at LEP and
the SLC. We focus our attention on genuinely anomalous quartic operators,
{\em i.e.} operators that do not modify the trilinear vertices. These
anomalous interactions cannot be constrained by the direct LEP II
analysis of the vertices $W^+ W^-\gamma$ and $W^+ W^-Z$. In general,
anomalous quartic couplings arise as the low-energy limit of heavy
state exchange, whereas trilinear couplings are modified by
integrating out heavy fields. Therefore, deviations on the triple
gauge-vector couplings should be harder to observe than the ones on
the quartic couplings, since the former are suppressed by factors of
$1/16 \pi^2$ \cite{aew}.  Furthermore, it is even possible to conceive
extensions of the SM where the trilinear couplings remain unchanged,
while the quartic vertices receive new contributions.  For instance,
the introduction of a new heavy scalar singlet, which interacts
strongly with the Higgs sector of the SM, enhances the quartic
vector-boson interaction without affecting either the triple
vector-boson couplings or the SM predictions for the $\rho$ parameter
\cite{singlet}.

At the one-loop level, anomalous quartic vector-boson interactions contribute
to the $Z$ pole physics through universal corrections to the gauge-boson
self-energies. In general, the oblique radiative corrections can be
parametrized in terms of three observables $S$, $T$, and $U$ \cite{pt}, or
equivalently $\epsilon^1$, $\epsilon^2$, and $\epsilon^3$ \cite{abc}. We shall
obtain the constraints on anomalous quartic vertices by imposing that their
one-loop contributions are compatible with the $Z$ pole data
\cite{altarelli,langacker}.  Since the SLC and LEP I achieved a precision of
the order of a per mille in some observables, the $Z$ pole physics is the best
available source of information on quartic vector-boson interactions.


\section{Theoretical Framework}

\medskip

If the electroweak symmetry breaking is due to a heavy (strongly
interacting) Higgs boson, which can be effectively removed from the
physical low-energy spectrum, or to no fundamental Higgs scalar at
all, one is led to consider the most general effective Lagrangian
which employs a nonlinear representation of the spontaneously broken
$SU(2)_L \otimes U(1)_Y$ gauge symmetry \cite{Appelquist}. The
resulting chiral Lagrangian is a non-renormalizable non-linear $\sigma$
model coupled in a gauge-invariant way to the Yang-Mills theory.  This
model independent approach incorporates by construction the low-energy
theorems \cite{cgg}, that predict the general behavior of Goldstone
boson amplitudes irrespective of the details of the symmetry breaking
mechanism. Notwithstanding, unitarity implies that this low-energy
effective theory should be valid up to some energy scale smaller than
$4\pi v \simeq 3$ TeV, where new physics would come into play.

To specify the effective Lagrangian one must first fix the symmetry
breaking pattern. We consider that the system presents a global
$SU(2)_L \otimes SU(2)_R$ symmetry that is broken to $SU(2)_C$. With
this choice, the building block\footnote{We follow the notation of
Ref.\ \cite{Appelquist}.} of the chiral Lagrangian is the
dimensionless unimodular matrix field $\Sigma(x)$, which transforms
under $SU(2)_L \otimes SU(2)_R$ as $(2,2)$:
\begin{equation}
\Sigma(x) ~=~ \exp\left(i \frac{\varphi^a(x) \tau^a}{v}\right) \; ,
\end{equation}
where the $\varphi^a$ fields are the would-be Goldstone fields and
$\tau^a$ ($a=1$, $2$, $3$) are the Pauli matrices.  The $SU(2)_L
\otimes U(1)_Y$ covariant derivative of $\Sigma$ is defined as
\begin{equation}
D_\mu \Sigma ~\equiv~ \partial_\mu \Sigma 
+ i g \frac{\tau^a}{2} W^a_\mu \Sigma -
i g^\prime \Sigma \frac{\tau^3}{2} B_\mu \; .
\end{equation}
At this point, it is convenient to introduce the following auxiliary
quantities
\begin{eqnarray}
T &\equiv& \Sigma \tau^3 \Sigma^\dagger
\; , \\
V_\mu &\equiv& \left ( D_\mu \Sigma \right ) \Sigma^\dagger
\; ,
\end{eqnarray}
which are $SU(2)_L$-covariant and $U(1)_Y$-invariant. Notice that $T$
is not invariant under $SU(2)_C$ custodial due to the appearance of
$\tau^3$ in its expression.

The lowest-order terms in the derivative expansion of the effective
Lagrangian are
\begin{equation}
{\cal L}^{(2)} = \frac{v^2}{4} \hbox{Tr} \left [ \left ( D_\mu \Sigma \right )
^\dagger \left ( D^\mu \Sigma \right ) \right ]
+ \beta_1 g'^2\frac{v^2}{4} \left ( \hbox{Tr}
\left [ T V_\mu \right ] \right )^2
\; .
\label{lagran2}
\end{equation}
The first term of the above equation is responsible for giving mass to
the gauge bosons $W^\pm$ and $Z$ for $ v = ( \sqrt{2} G_F )^{-1}
$. The second term violates the custodial $SU(2)_C$ symmetry and
contributes to $\Delta\rho$ at the tree level, being strongly
constrained by the low-energy data. This term can be understood as the
low-energy remnant of the high-energy custodial symmetry breaking
physics, which has been integrated out above a certain scale
$\Lambda$.  Moreover, at the one-loop order, it is also required in
order to cancel the divergences in $\Delta\rho$, arising from diagrams
containing a hypercharge boson in the loop \cite{Appelquist}.  This
subtraction renders $\Delta\rho$ finite, although dependent on the
renormalization scale.

At the next order in the derivative expansion $D=4$, there are many
operators that can be written down \cite{Appelquist}. We shall
restrict ourselves to the ones that exhibit genuine quartic
vector-boson interactions. These operators are
\begin{eqnarray}
{\cal L}^{(4)}_4 &=& \alpha_4\left[{\rm{Tr}}
\left(V_{\mu}V_{\nu}\right)\right]^2
\label{eff:4}
\;, \\
{\cal L}^{(4)}_5 &=& \alpha_5\left[{\rm{Tr}}
\left(V_{\mu}V^{\mu}\right)\right]^2
\;, \\
{\cal L}^{(4)}_6 &=& \alpha_6 \; {\rm{Tr}}\left(V_{\mu}V_{\nu}\right)
{\rm{Tr}}
\left(TV^{\mu}\right){\rm{Tr}}\left(TV^{\nu}\right) \;, \\
{\cal L}^{(4)}_7 &=& \alpha_7\;{\rm{Tr}}\left(V_{\mu}V^{\mu}\right)
\left[{\rm{Tr}}\left(TV^{\nu}\right)\right]^2
\;, \\
{\cal L}^{(4)}_{10} &=& \alpha_{10}\left[{\rm{Tr}}\left(TV_{\mu}\right)
\;{\rm{Tr}}\left(TV_{\nu}\right)\right]^2
\; .
\label{eff:10}
\end{eqnarray}
In an arbitrary gauge, these Lagrangian densities lead to quartic
vertices involving gauge bosons and/or Goldstone bosons. In the
unitary gauge, these effective operators give rise to anomalous $ZZZZ$
(all operators), $W^+W^-ZZ$ (all operators except ${\cal
L}^{(4)}_{10}$), and $W^+ W^- W^+ W^-$ (${\cal L}^{(4)}_4$ and ${\cal
L}^{(4)}_5$) interactions.  Moreover, the interaction Lagrangians
${\cal L}^{(4)}_6$, ${\cal L}^{(4)}_7$, and ${\cal L}^{(4)}_{10}$
violate the $SU(2)_C$ custodial symmetry.  Notice that quartic
couplings involving photons remain untouched by the genuinely quartic
anomalous interactions at the order $D=4$.  The Feynman rules for the
quartic couplings generated by these operators can be found in the
last article of Ref.\ \cite{Appelquist}.

In our calculations, we adopted an arbitrary $R_\xi$ gauge, whose 
gauge-fixing Lagrangian is
\begin{equation}
{\cal L}_{GF} = - \frac{1}{2 \xi_B} f_0^2 -  \frac{1}{2\xi_W}
\left ( \sum_{i=1}^3 f_i^2 \right )
\; ,
\nonumber
\end{equation}
where
\begin{eqnarray}
f_0 &=& \partial_\mu B^\mu - \frac{i}{4} g^\prime v \xi_B
\hbox{Tr} ( \tau^3 \Sigma) \; ,
\\
f_i &=& \partial_\mu W_i^\mu + \frac{i}{4} g v \xi_W \hbox{Tr}
(\tau^i \Sigma) \; ,
\end{eqnarray}
with $g$ ($g^\prime$) being the $SU(2)_L$ ($U(1)_Y$) coupling constant.

At the one-loop level, the effective interactions (\ref{eff:4}) --
(\ref{eff:10}) contribute to the $Z$ physics only through corrections
to the gauge boson propagators ($\Sigma_{new}$).  The anomalous
oblique corrections can be efficiently summarized in terms of the
parameters $S_{new}$, $T_{new}$, and $U_{new}$ \cite{pt}, or the
equivalent set $\epsilon^1_{new}$, $\epsilon^2_{new}$, and
$\epsilon^3_{new}$ \cite{abc}, whose expressions as functions of the
unrenormalized gauge boson self-energies are
\begin{eqnarray}
\frac{\alpha S_{new}}{4 s_w^2} &\equiv& \frac{1}{M_Z^2} \left \{
c_W^2 \left [  \Sigma_{new}^\gamma(M_Z^2) + \Sigma_{new}^Z(0)
- \Sigma_{new}(M_Z^2)\right ]
\right.
\nonumber \\
&& \left. - s_W c_W \left ( \frac{c_W^2}{s_W^2} - 1 \right )
\left [ \Sigma_{new}^{\gamma Z}(M_Z^2) - \Sigma_{new}^{\gamma Z}(0)
\right ] \right \}~=~\epsilon^3_{new}
\; , \\
\alpha T_{new} &\equiv& \frac{\Sigma_{new}^Z(0)}{M_Z^2}
- \frac{\Sigma_{new}^W(0)}{M_W^2}
-2 \frac{s_W}{c_W} \frac{\Sigma_{new}^{\gamma Z}(0)}{M_Z^2}=\epsilon^1_{new}
\; , \\
\frac{\alpha U_{new}}{4 s_W^2} &\equiv&
\left \{
\frac{\Sigma_{new}^W(0) -  \Sigma_{new}^W(M_W^2)}{M_W^2} +
s_W^2 \frac{\Sigma_{new}^\gamma(M_Z^2)}{M_Z^2}
\right.
\nonumber \\
&& \left. - 2 s_W c_W \frac{\Sigma_{new}^{\gamma Z}(M_Z^2)
- \Sigma_{new}^{\gamma Z}(0)}{M_Z^2}
+ c_W^2 \frac{\Sigma_{new}^Z(M_Z^2) - \Sigma_{new}^Z(0) }{M_Z^2}
\right \}=-\epsilon^2_{new}
\; ,
\end{eqnarray}
where $\alpha$ is the fine structure constant and $s_W$ ($c_W$) is the
sine (cosine) of the weak mixing angle. These expressions for
$S_{new}$, $T_{new}$, and $U_{new}$ are valid for an arbitrary
momentum dependence of the vacuum polarization diagrams \cite{kk}; 
they recover the original definitions of Ref.\ \cite{pt}, when we
consider only the first two terms in the momentum expansion of the
self-energies. This is the case of the present work.  Since the
contribution from the new operators to the Z observables occurs only
through the gauge-boson vacuum polarization diagrams, which are
momentum-independent, we can also express the new contributions in
terms of the $\epsilon$ parameters from Ref.\ \cite{abc}.

Recent global analyses of the LEP, SLD, and low-energy data yield the
following values for the oblique parameters \cite{altarelli}:
\begin{equation}
\begin{array}{l}
\epsilon^1 ~=~ \epsilon^1_{SM}+\epsilon^1_{new}
~=~ (5.1\pm 2.2)\times 10^{-3}
\; , \\
\epsilon^2 ~=~ \epsilon^2_{SM}+\epsilon^2_{new}
~=~ (-4.1\pm 4.8)\times 10^{-3}
\; , \\
\epsilon^3 ~=~ \epsilon^3_{SM}+\epsilon^3_{new}
~=~ (5.1\pm 2.0)\times 10^{-3}
\; . \\
\end{array}
\label{epsilons}
\end{equation}
In Ref.\ \cite{langacker}, the results are obtained in terms of
$S_{new}$, $T_{new}$, and $U_{new}$, consistent with Eqs.\
(\ref{epsilons}). In order to extract the value of the oblique
parameters due to new physics, we must subtract the SM contribution,
which depends upon the SM parameters, in particular, on the top quark
mass $m_{top}$.


\section{Results and Conclusions}

\medskip

We used dimensional regularization \cite{dim} to evaluate the one-loop
 contributions from the effective interactions (\ref{eff:4}) --
 (\ref{eff:10}), in order to preserve gauge invariance and to keep the
 ordering of the different contributions simple.  In analogy to what
 happens for chiral Lagrangians applied to low-energy QCD \cite{gl},
 the electroweak chiral Lagrangian leads to an expansion in powers of
 the momentum $p$ and the weak coupling constant $g$. This $g$
 dependence is due to the introduction of new degrees of freedom
 associated to the gauge bosons.  At a given order in $g$ and $p$,
 there are just a finite number of operators and loop diagrams that
 contribute to a process. Therefore, the effective theory
 renormalization can be carried out by renormalizing the coupling
 constants of the operators that appear in the process at the order
 that the analysis is being done. In this work, we evaluate the
 one-loop contributions from $D=4$ interactions, where $D$ counts the
 number of derivative plus the numbers of gauge bosons in the
 operator, which lead to $g^4 p^2$ corrections. In a complete
 calculation, we should also include the effects of two-loop graphs of
 $D=2$ operators and tree-level contributions from $D=6$ operators.
 At the end of the day, the experimental results for the oblique
 parameters would constrain combinations of the coupling constants
 appearing in the $D=2$, $D=4$, and $D=6$ operators
 \cite{gl}. However, this next-to-next-to-leading order calculation
 contains a large number of free parameters that reduces its
 usefulness.  Notwithstanding, we can bound the anomalous quartic
 interactions from their oblique corrections under the naturalness
 assumption that no cancelation takes place amongst the ${\cal
 L}^{(2)}$, ${\cal L}^{(4)}$, and ${\cal L}^{(6)}$ contributions that
 appear at the same order in the expansion.  The motivation for this
 assumption is that the effects of the new physics and states from
 higher-energy scales must manifest themselves in a very clear way,
 otherwise they are very hard to observe.

Our procedure to bound the operators (\ref{eff:4}) -- (\ref{eff:10}) is
the following: first we evaluate their oblique corrections using
dimensional regularization. Then, we use the leading non-analytic
contributions from the loop diagrams to constrain the quartic
interactions -- that is, we keep only the terms proportional to
log$(\mu^2)$, dropping all others. The contributions that are relevant
for our analysis are easily obtained by the substitution
\begin{equation}
\frac{2}{4-d} \rightarrow {\rm{log}}\;\frac{\Lambda^2}{M_Z^2}\; ,
\nonumber
\end{equation}
where $\Lambda$ is the energy scale which characterizes the appearance
of new physics.

Using the above procedure, we obtained that $S_{new} = U_{new} = 0$
and that only $T_{new}$ is non-vanishing for all the quartic anomalous
interactions, being given by
\begin{eqnarray}
\alpha T_{new}& =\epsilon^1_{new}= & - \frac{15 \alpha_4}{64 \pi^2}g^4
(1+c_W^2)\frac{s_W^2}{c_W^2}
\; {\rm{log}}\;\frac{\Lambda^2}{M_Z^2}
\; , \\
\alpha T_{new} & = \epsilon^1_{new}= & - \frac{3 \alpha_5}{32 \pi^2}g^4
(1+c_W^2)\frac{s_W^2}{c_W^2}
{\rm{log}}\;\frac{\Lambda^2}{M_Z^2}
\; , \\
\alpha T_{new} &=  \epsilon^1_{new}=&  - \frac{3 \alpha_6}{64 \pi^2} g^4
\left ( 2 + \frac{11}{c_W^4}
\right ) {\rm{log}}\;\frac{\Lambda^2}{M_Z^2}
\; , \\
\alpha T_{new} &=  \epsilon^1_{new}=&  - \frac{3 \alpha_7}{64 \pi^2} g^4
\left( \frac{1+c_W^4}{c_W^4}\right)
{\rm{log}}\;\frac{\Lambda^2}{M_Z^2}
\; , \\
\alpha T_{new} &=  \epsilon^1_{new} =&  -\frac{9 \alpha_{10}}{8 \pi^2} g^4
\left(\frac{1}{c_W^4}\right)
{\rm{log}}\;\frac{\Lambda^2}{M_Z^2}
\; ,
\end{eqnarray}
for ${\cal L}^{(4)}_4$, ${\cal L}^{(4)}_5$, ${\cal L}^{(4)}_6$,
  ${\cal L}^{(4)}_7$ e ${\cal L}^{(4)}_{10}$, respectively.

Our calculation has been done in a general $R_\xi$ gauge and we have
explicitly verified the cancellation of the $\xi$-dependent terms,
indicating that our result is gauge-invariant. Ward identities relate
the two- and three-point functions, and consequently the anomalous
contributions to the two-point functions are gauge-independent since
there is no one-loop three-point contribution due to the effective
quartic interactions.

Our first step towards obtaining the bounds on the anomalous quartic
vertices is to determine the SM contribution to $\epsilon^1$. As
discussed above, the gauge-boson contribution to this parameter is
infinite as a consequence of the absence of the elementary Higgs. On
the other hand, we must also include the tree level effect due to the
$\beta_1$ operator in Eq.\ (\ref{lagran2}), which absorbs this infinity
through the renormalization of the $\beta_1$ constant.  If the
renormalization condition is imposed at a scale $\Lambda$, we are left
with the contribution due to the running of $\beta_1$ from the scale
$\Lambda$ to $M_Z$. Therefore, the SM contribution without the Higgs boson
will be the same as that of the SM with an elementary Higgs, with the
substitution $\ln (M_H)\rightarrow \ln(\Lambda)$.

We show in Table \ref{limits} the 90\% CL constraints on the quartic
anomalous vector-boson interactions which are obtained from Eq.\
(\ref{epsilons}) assuming that $\Lambda = 2$ TeV. In this case
the SM contribution to $\epsilon^1$ is in the range
($2.68$--$7.58$)$\times 10^{-3}$ for $m_{top}=170$--$220$ GeV.  Our
bounds for $\alpha_4$ and $\alpha_5$ agree with the ones in Ref.\
\cite{dv} which were obtained in the unitary gauge.

It is interesting to notice that our analysis does not show any
indication of new physics beyond the SM since all the anomalous
couplings are compatible with zero at 90\% CL. A natural order of
magnitude of the anomalous couplings $\alpha_i$ in a fundamental gauge
theory is $ g^2 v^2/\Lambda^2$ \cite{aew}, since the quartic anomalous
interactions can be generated by tree diagrams. Thus, we might expect
that the size of the $\alpha$'s should be of the order of
$M_Z^2/\Lambda^2 \simeq 2 \times 10^{-3}$. From our results we see
that only the operators that break the custodial $SU(2)_C$ symmetry,
(${\cal L}^{(4)}_{6,7,10}$) get limits close to this
expectation\footnote{ It is interesting to notice that models
containing spin-0 and spin-1 resonances also lead to couplings of this
order \cite{bdv}.}.

Future colliders will be able to search for anomalous quartic
interactions through multiple gauge-boson production. 
Assuming, as in Ref.\ \cite{bdv}, that an anomalous coupling is
observable at the LHC if it induces a 50\% change in the integrated
cross section for the production of pairs $V_L V_L$ ($V=W^\pm$, $Z$),
it will be possible to detect the couplings $\alpha_4$ and $\alpha_5$
provided they satisfy $|\alpha_4,\alpha_5|\sim {\cal O}(0.005)$
\footnote{These limits should be taken with a grain of salt since the
analysis in Ref.\ \cite{bdv} does not take into account the efficiency
to detect the gauge bosons. This will probably make the limits be a
factor of 10 weaker.}.  These constraints are stronger than the limits
obtained from the $Z$ physics, as also concluded in Ref.\
\cite{dv}. In Ref.\ \cite{bela1}, 
the capabilities of the NLC to study quartic anomalous couplings via
the production mechanisms $e^+ e^- \rightarrow W^+ \, W^-\, Z$ and
$e^+ e^- \rightarrow Z \, Z \, Z$ are analysed, this last mechanism
being the one yielding stronger constraints.  Using their values for
the $ZZZ$ cross section associated with the couplings $\alpha_{4,5}$,
we translated their results into 90\% CL limits $|\alpha_4, \alpha_5|
\lesssim 0.2$, $|\alpha_6, \alpha_7|
\lesssim 0.1$, and $|\alpha_{10}| \lesssim 0.05$. These future
direct bounds are weaker than the limits already imposed by the LEP
data for most of the quartic anomalous couplings. Therefore, we can
see that our results show that only the LHC can improve what we have
learned from the radiative corrections at the $Z$ pole.

Summarizing, we have analyzed the effects of possible anomalous
quartic vector-boson interactions that appear in a scenario where
there is no particle associated to the symmetry-breaking sector in
the low-energy spectrum. Using a chiral Lagrangian at the order $D=4$
and an arbitrary gauge $R_\xi$, we draw the limits on the anomalous
interactions $ZZZZ$, $W^+W^-ZZ$, and $W^+W^-W^+W^-$ arising from the
precision measurements at the $Z$ pole. We extended previous results
by considering all possible anomalous couplings that appear at order
$D=4$ and by working in an arbitrary $R_\xi$ gauge. Our analysis shows
that, with the present limits, these anomalous couplings are already
out of reach of a 500 GeV $e^+e^-$ collider for most of the values of
$m_{top}$. However, the LHC will be able to further extend the bounds
on some of these couplings.


{\bf\large Acknowledgements}

This work was partially supported by the U.S.\ Department of Energy
under Grants Nos.\ DE-FG02-95ER40896 and DE-FG02-91ER40661, by the
University of Wisconsin Research Committee with funds granted by the
Wisconsin Alumni Research Foundation, by Conselho Nacional de
Desenvolvimento Cient\'{\i}fico e Tecnol\'ogico (CNPq), and by
Funda\c{c}\~ao de Amparo \`a Pesquisa do Estado de S\~ao Paulo
(FAPESP).


\begin{table}
\caption{Limits on the anomalous quartic vector boson couplings at 90\% CL for
$\Lambda=2$ TeV.}
\label{limits}
\begin{displaymath}
\begin{array}{|c|c|c|}
\hline
m_{top}=170~ \mbox{GeV} &  m_{top}=200~ \mbox{GeV} &
m_{top}=220~ \mbox{GeV} \\
\hline
 -0.060\leq \alpha_4\leq 0.30 & -0.20\leq \alpha_4\leq 0.16 &
-0.30\leq \alpha_4\leq 0.056\\ \hline
 -0.15\leq \alpha_5\leq 0.76 & -0.50\leq \alpha_5\leq 0.40 &
-0.77\leq \alpha_5\leq 0.14\\ \hline
 -0.010\leq \alpha_6\leq 0.053 & -0.035\leq \alpha_6\leq 0.028 &
-0.054\leq \alpha_6\leq 0.0099 \\ \hline
 -0.077\leq \alpha_7\leq 0.39 &  -0.26\leq \alpha_7\leq 0.21 &
-0.39\leq \alpha_7\leq 0.072\\ \hline
 -0.0051\leq \alpha_{10}\leq 0.026 &  -0.017\leq \alpha_{10}\leq 0.014 &
-0.026\leq \alpha_{10}\leq 0.0048 \\ \hline
\end{array}
\end{displaymath}
\end{table}


\begin{figure}
\epsfxsize=10cm
\begin{center}
\mbox{\epsfig{file=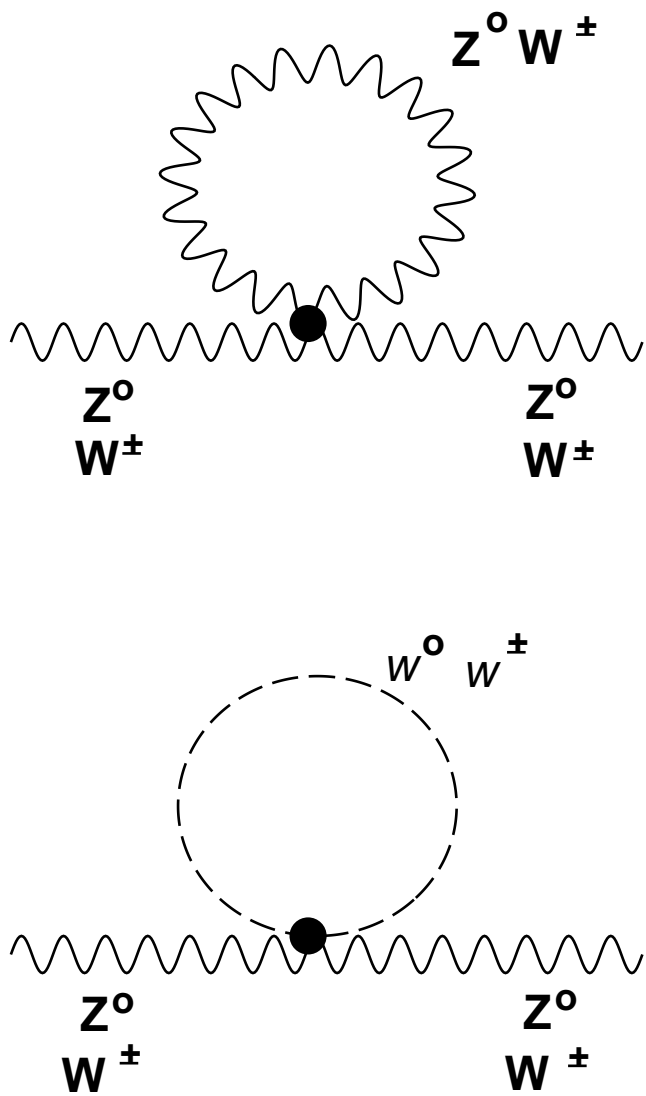,bbllx=60pt,bblly=200pt,bburx=600pt,%
bbury=600pt}}
\end{center}
\caption{Feynman diagrams that contain contributions from anomalous
quartic vertices.}
\label{diagram}
\end{figure}

\end{document}